# AMPCliff: quantitative definition and benchmarking of activity cliffs in antimicrobial peptides

Kewei Li[1], Yuqian Wu[2], Yinheng Li[4], Yutong Guo[3], Yan Wang[7], Yiyang Liang[8], Yusi Fan[1], Lan Huang[1], Ruochi Zhang[5,*], Fengfeng Zhou[1,6,*].

1 College of Computer Science and Technology, and Key Laboratory of Symbolic Computation and Knowledge Engineering of Ministry of Education, Jilin University, Changchun, Jilin, China, 130012.

2 School of Software, Jilin University, Changchun 130012, Jilin, China.

3. School of Life Sciences, Jilin University, Changchun 130012, Jilin, China.

4 Department of Computer Science, Columbia University, 116th and Broadway, New York City, New York, 10027, United States.

5 School of Artificial Intelligence, and Key Laboratory of Symbolic Computation and Knowledge Engineering of Ministry of Education, Jilin University, Changchun, Jilin, China, 130012.

6 School of Biology and Engineering, Guizhou Medical University, Guiyang 550025, Guizhou, China.

7 School of Computer Engineering, Changchun University of Engineering, Changchun 130103, Jilin, China.

8 Changchun Wenli High School, Changchun 130062, Jilin, China.

* Correspondence may be addressed to Fengfeng Zhou: FengfengZhou@gmail.com or ffzhou@jlu.edu.cn. Correspondence may also be addressed to Ruochi Zhang: zrc720@gmail.com.

Emails: Kewei Li, kwbb1997@gmail.com; Yuqian Wu, 924476388@qq.com; Yutong Guo, 3291865187.qq.com; Yinheng Li, yl4039@columbia.edu; Yusi Fan, fan_yusi@163.com; Ruochi Zhang, zrc720@gmail.com; Lan Huang, huanglan@jlu.edu.cn; Fengfeng Zhou, FengfengZhou@gmail.com.

## Abstract

Since the mechanism of action of drug molecules in the human body is difficult to reproduce in the in vitro environment, it becomes difficult to reveal the causes of the activity cliff phenomenon of drug molecules. We found out the AC of small molecules has been extensively investigated but limited knowledge is accumulated about the AC phenomenon in peptides with canonical amino acids. Understanding the mechanism of AC in canonical amino acids might help understand the one in drug molecules. This study introduces a quantitative definition and benchmarking framework AMPCliff for the AC phenomenon in antimicrobial peptides (AMPs) composed by canonical amino acids. A comprehensive analysis of the existing AMP dataset reveals a significant prevalence of AC within AMPs. AMPCliff quantifies the activities of AMPs by the MIC, and defines 0.9 as the minimum threshold for the normalized BLOSUM62 similarity score between a pair of aligned peptides with at least two-fold MIC changes. This study establishes a benchmark dataset of paired AMPs in *Staphylococcus aureus* from the publicly available AMP dataset GRAMPA, and conducts a rigorous procedure to evaluate various AMP AC prediction models, including nine machine learning, four deep learning algorithms, four masked language models, and four generative language models. Our analysis reveals that these models are capable of detecting AMP AC events and the pre-trained protein language model ESM2 demonstrates superior performance across the evaluations. The predictive performance of AMP activity cliffs remains to be further improved, considering that ESM2 with 33 layers only achieves the Spearman correlation coefficient 0.4669 for the regression task of the MIC values on the benchmark dataset. Source code and additional resources are available at https://www.healthinformaticslab.org/supp/ or https://github.com/Kewei2023/AMPCliff-generation.

## Introduction

Activity cliff (AC) refers to a pair of similar compounds that differ by only a small structural change but exhibit a large difference in their biochemical activities [1-11]. AC was firstly observed in the investigations of Quantitative Structure-Activity Relationship (QSAR) modeling in drug design [12], and remains as a persistent challenge in QSAR modeling for over 30 years [1]. This phenomenon affects various tasks ranging from virtual screening to lead optimization in drug development [9,13]. Precise AC modeling is highly beneficial for pharmaceutical researchers.

Advances in deep learning has significantly driven the development of drug design and the applications of QSAR modeling in both scientific research and industry [14-16]. Deep learning methods face many challenges, including interpretability, data quality dependency, and generalization. These challenges are particularly evident in handling ACs in drug design [1,10,15,17,18]. Deng et al., detected that nearly half of the molecular pairs in the Multi-Drug Resistance Protein 1 (MDR1) dataset are ACs, and the Root Mean Square Error (RMSE) for the regression task of the AC pairs is increased by at least 0.6 using the Graph Convolutional Network (GCN) model compared to non-AC pairs [17].

Recent efforts in QSAR modeling of antimicrobial peptides (AMP) include Recurrent Neural Network (RNN) models [19] and generative variational autoencoder (VAE) [20]. Despite these advancements, the best models have only achieved a Pearson correlation coefficient (PCC) of 0.77. This observation suggests that ACs could be a critical factor in building precise QSAR models for AMPs.

The potential AC phenomenon in AMPs was discovered at least 15 years ago. In 2008, Koo et al. investigated the structure-activity relationship of Parasin I, a 19-amino acid histone H2A-derived antimicrobial peptide that killed S. aureus by cell membrane [21]. They found that the deletion of one lysine residue in the N termini resulted in loss of antimicrobial activity, but did not affect the α-helical content of the peptide. The antimicrobial activity was recovered when the lysine residue was substituted with another basic residue, arginine, but not with polar, neutral, or acidic residues [21]. But there were limited results directly quantitative defined the AC phenomenon in AMPs with canonical amino acids. Since canonical amino acids have more information than molecules, the definition of AC in canonical amino acids should be different from the one in molecules.

AC studies in small molecules can be separated into three types (Fig. 1): (i) discovery, (ii) data-driven analysis, and (iii) predictive modeling. Since 2006, the concept of AC received multiple precise definitions [3-5,9,22], such as MMP-Cliff [3] and 3D activity cliffs [5]. Various metrics [2,7,23] like SARI [23] and SALI [2] have also been proposed to characterize ACs. The term MMP-Cliff was defined to focus on structural differences within molecule pairs in 2012 [3], and remains the most prevalent form in the AC characterizations. The same team also explored 3D activity cliffs [5], which highlighted the challenges of applying 2D measures to 3D structures and emphasized the importance of molecular representation in AC characterization. Despite such advances, the exploration of 3D cliffs has stalled since 2015 [8], possibly due to limited data. The popularly-used term MMP-Cliff requires two molecules in the MMP (Matched Molecular Pair) to differ at most eight non-hydrogen atoms in the exchanged fragments of the size at most 13 non-hydrogen atoms, and to have the potency difference at least 2 orders of magnitudes.

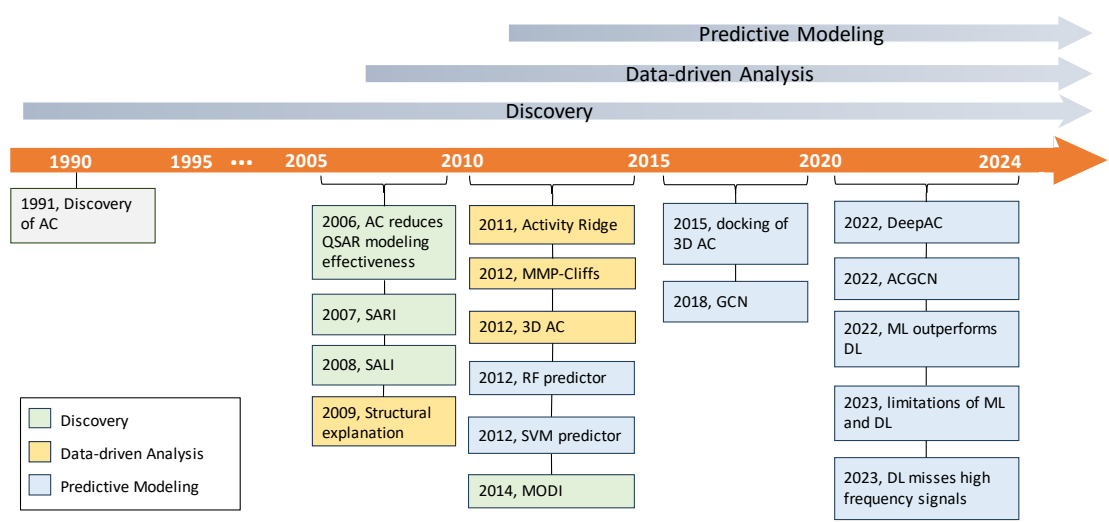

**Fig. 1 A brief history of the investigations of the AC phenomenon for molecular prediction task.**

Most AC studies before 2010 focused on the data-drive analysis of what caused the AC phenomenon (Fig. 1). Afterwards, numerous predictive models based on the definition of MMP-Cliff have been developed [1,9], e.g., random forests [24] and support vector machines [25]. Deep learning (DL) algorithms have also been explored for their applications in this field and delivered limited successes [1,6] due to their inability to capture the high frequency signals in the AC pairs [18,26-29]. Machine learning (ML) models with stereochemistry-aware features [6] like Extended Connectivity Fingerprints (ECFP) have proven more effective in characterizing the MMP-Cliff molecular pairs than end-to-end deep learning algorithms [6,11]. Recent research has taken a fresh look at QSAR models from an algorithmic perspective, and pinpointed fundamental issues with deep learning in predicting small molecule properties, beyond just data scarcity [18].

This study quantitatively defines the AC Phenomenon in AMP as a pair of AMPs with high sequence similarity but radically different antimicrobial activities (called an AMPCliff). We systematically evaluate the recent peptide representation algorithms on the AMPCliff prediction task. To the best of our knowledge, there has no direct mention of ACs in AMPs with canonical amino acids. We reviewed the history of AC investigations in small molecules since its concept has been extensively studied. We identify that the popularly-employed definition of MMP-Cliff in small molecules fails to consider the evolutionary conservations among peptides. It is widely accepted that two proteins or peptides with evolutionarily conserved substitutions tend to exert similar biochemical activities [30]. This oversight is understandable that evolutionary conservation is not as important in small molecules as in peptides. We hope understanding the mechanism of AC in AMP can help understand the one in drug molecules in vitro environment.

The prevalence of AMPs with canonical amino acids in the public datasets constitutes the basis of quantitatively defining and benchmarking the AMPCliff pairs in this study. The main contributions of this study are summarized as follows:

- We present the first systematic screening of the AC Phenomenon in AMPs (AMPCliff), based on the quantitative activity measurement of their Minimum Inhibitory Concentrations (MICs) against the same bacterium.

- The quantitative definition of AMPCliff is proposed based on the AMPs of S. aureus in the public AMP dataset GRAMPA. we propose a novel data partition method, AC split as a benchmark data split strategy for model training.

- We conduct a comprehensive benchmarking experiment of various machine learning, deep learning, and pre-trained language models on the AMPCliff prediction task.

## Preliminaries

### Peptide representations

**Fixed type.** Over the years, various algorithms have been proposed to represent peptide sequences [31-34], including (i) physicochemical properties like molecular weight and isoelectric point [32], (ii) compositional components like amino acid composition (AAC) and pseudo AAC (Pse-AAC) [33], and (iii) structural features [31] like α-helix (AHP) propensities [34].

These fixed type representation features were well defined in the literature and are also called hand-crafted features [20,35]. The fixed type representations can even outperform deep learning-based representations in some classification tasks [36]. Garcia-Jacas et al., further demonstrated that the fixed type features usually contributed complementary information to the deep learning-based representations and their combinations tended to achieve the best prediction performances [36]. Based on these findings, traditional fixed type features continue to be employed in peptide property prediction tasks. Following the approach described in the literature [19], this study calculates the following fixed types of features, including molecular weight, charge characteristics, content of polar and nonpolar amino acids, hydrophobicity, and van der Waals volume. We also include amino acid composition (AAC), dipeptide composition (DiC), pseudo AAC (Pse-AAC), amphipathic Pse-AAC (APAAC), Moreau-Broto Autocorrelation, and Quasi-Sequence Order. The structural feature α-helix propensity is also employed in this study. The total list of 677 fixed type representation features can be found in the Supplementary Table S.

**Dictionary index.** This method assigns sequential integers starting from 1 to each amino acid, and it is widely used in the recent studies of peptide design [19].

**One-hot encoding.** This popularly-used peptide feature representation method converts each amino acid's numerical identifiers into vectors with only 0 and 1. For example, if the dictionary size is *n*, the feature vector for the $k^{th}$ amino acid will be <$f(i)$> ($i \in \{1, 2, ..., n\}$), where $f(i)=1$ if $i=k$, and $f(i)=0$ if $i \neq k$.

However, one-hot encoding does not incorporate the correlations between the physical and chemical properties of amino acids. Similar to the dictionary index, this method allows the model to memorize the occurrence frequencies of amino acids. Another limitation of one-hot encoding is its inability to index out-of-distribution (OOD) amino acids. If an amino acid that has not been seen in the training set appears in the test set, one-hot encoding will not be able to represent them.

**Word embedding.** Researchers employed the idea of word embedding from the area of Natural Language Processing (NLP) to let the model learn a general representation of amino acids. The success of BERT and GPT models from the NLP area leads researchers to regard peptide and protein sequences as the "language of life" and employ the word embedding method to represent these sequences [37,38].

The word embedding method has the similar limitation as one-hot encoding, i.e., inability of representing OOD amino acids. However, this would not be a problem for peptides and proteins with canonical amino acids. This study focuses on the AMPs composed of canonical amino acids, which make up a vocabulary of 20 distinct residues.

**Fingerprints.** This method was originally developed for the representations of molecules, which have smaller molecular weights than peptides [14]. One of the popularly-used fingerprint method for molecules is the Extended Connectivity Fingerprint (ECFP) [39]. A fingerprint method generates a unique representation of a molecule by taking into account the inter-atom connectivity in the molecule, and is widely used in compound similarity searches and drug discovery studies [40]. This method only requires the SMILES (Simplified Molecular Input Line Entry System) expression of a molecule for calculation, and can also describe peptides with both canonical and non-canonical amino acids. Although not specifically designed for peptides with canonical amino acids, the fingerprint method can also represent a peptide if the SMILES expressions of the canonical amino acids are connected in their order in the peptide (see **Fig. 3e**) [40].

Most of the existing pre-trained peptide representation models are based on word embedding method probably due to the simplicity in the expression sequence of canonical amino acids.

**Model architectures**

Various model architectures have been proposed for peptide property prediction, such as Recurrent Neural Network (RNN) [19,20], one-dimensional convolutional neural network (1d-CNN) [20,40], and transformers [37,38,41]. RNN was designed for handling sequential data (e.g., text and audio), and can be easily employed to work on peptides represented as sequences of amino-acid-level features like dictionary index, and one-hot encoding. The 1d-CNN was originally designed for handling images, and is good at capturing local features of peptide sequences. Self-supervised learning (SSL) has been recently proposed for pre-training protein representation models on large-scale unlabeled protein corpus to address the scarcity of annotated data in drug discovery before the downstream fine-tuning [37,38,41].

This study employs two types of pre-trained models: transformer encoder-based models like BERT [42] and ESM2 [41] with 6, 12, 33 layers, and transformer decoder-based models like GPT2 [43] and the small/base/medium versions of ProGen2 [44]. All these pre-trained models use peptide sequences as input. To evaluate the effectiveness of these pre-trained peptide representation models, we use traditional machine learning models on the fixed types of representations as baselines.

**Traditional ML models:** ML algorithms have been successfully employed in the peptide property prediction tasks. Random forest (RF) is an ensemble of decision tree predictors. Guha et al. presented one of the first few models to predict AC property [24]. RF has also been widely employed in drug discovery prior to the "deep-learning" era [45]. Gradient Boosting (GB) is another tree-based algorithm used for AC property prediction [10]. GB iteratively trains a series of weak tree models to minimize a loss function and tries to add new models to correct the overall errors. Tamura et al., applied the overfitting-resistant and parallelizable ensemble learning model XGBoost (eXtreme Gradient Boosting) to the AC property prediction task [46].

Gaussian Process (GP) is a probabilistic model where predictions at any set of points are assumed to have a joint Gaussian distribution. It calculates prediction results of estimated uncertainty for both regression and classification tasks, has been also successfully applied for AC prediction task [47,48]. Support Vector Machine (SVM) is a classical supervised algorithm for both classification and regression tasks [49]. It tries to find a hyperplane that separates different classes in a dataset with the maximized margins. AC property prediction is just one of the successful applications of SVM [25].

**DL models:** RNN is a class of neural network where connections between nodes form a directed graph along a temporal sequence, and can capture temporal dynamic behaviors within the input sequential data [50]. The long short-term memory (LSTM) was a variant of RNN with a special focus on the long-term dependencies, and was successfully applied to peptide design based on dictionary index[19] and one-hot encoding[20] features, respectively.

Convolutional neural network (CNN) is widely used in extracting imagery patterns through its layers of convolutional filters, and the cascaded convolutional layers are efficient in capturing spatial hierarchies within images. Recent studies engineered peptide sequences as their one-hot encoding [20] and fingerprint [40] representations for the CNN-based peptide design tasks.

**Pre-trained Language models:** Protein/peptide sequences can be viewed as "the language of life" [37], and the pre-trained language models from the NLP area have been successfully transferred to represent protein/peptide sequences. The Bidirectional Encoder Representations from Transformers (BERT) framework was pre-trained by large English text corpus, and its training paradigm closely aligns with the characteristics of proteins/peptides [42]. The observations lead to the pre-training of the ESM2 [41], whose performance gets better as its number of hyper-parameters increases.

NLP researchers presented many applications of the transformer decoder-based model Generative Pre-Trained Transformer (GPT) pre-trained by English corpus [51], whose potential scaling capacity has also been evaluated on proteins [44,52]. Proteins and peptides are both chains of amino acids, and proteins usually fold into specific shapes to exert their biological functions. Peptides are much shorter than proteins, and usually carry 50 or fewer amino acids in general. So it is interesting to explore the capacity of these pre-trained language models on the AC property prediction task.

**Benchmark dataset**

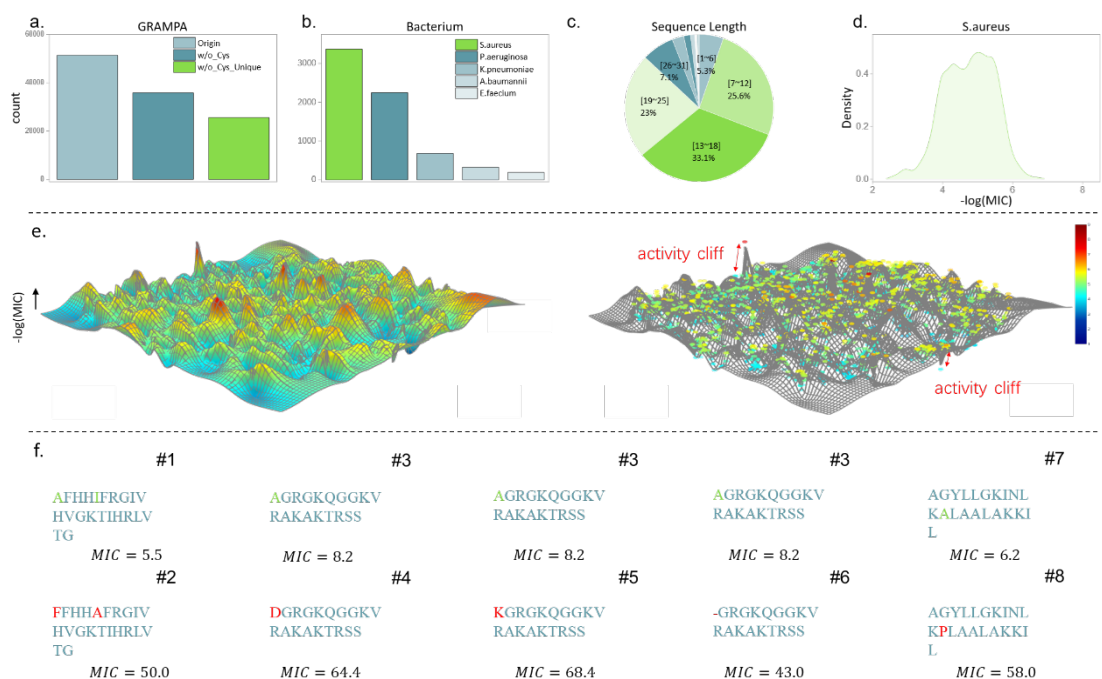

**Fig. 2 Exploratory profiling for GRAMPA dataset. a** The number of filtered peptides by excluding sequences with Cysteine (w/o_Cys) and dropping duplicate sequences (w/o_Cys_Unique) from the original GRAMPA dataset (Origin). **b** The number of peptides against a bacterium after filtering. *S. aureus, P. aeruginosa, K. pneumoniae, A. baumannii, E. faecium* are the 5 out of the 6 pathogens that pose the greatest threat to human life. **c** Length distribution of peptides against *S. aureus*. **d** Label distribution of peptides against *S. aureus*. **e** The landscape of the peptides again *S. aureus*. Features were the last layer of ESM2 with 33 layers reduced by t-SNE. **f** ACs showcased on series of peptides with 5-fold MIC changes. The unit of MIC is $\mu M$, and the original data are available in the GRAMPA dataset.

**Description of the benchmark dataset.** Building a QSAR model for AMP design has been significantly hindered due to the limited availability of quantitative measurements of AMP activity, such as MIC values, in public datasets like APD3 [53], DRAMP [54], DBAASP [55], and YADAMP [56]. Witten et al. finally curated the MIC values of AMPs from these public datasets in the Spring of 2018 [57]. To the best of our knowledge, GRAMPA [57] is the first and only public AMP dataset with MIC values, and it has already facilitated the studies of peptide design. Huang et al. trained an RNN regressive model based on the GRAMPA dataset to filter highly active peptides [19]. Pandi et al. also filtered sequences against E.coli and B. subtilis from GRAMPA to build RNN-based and CNN-based regressive models [20]. GRAMPA contains 6760 unique sequences, and

51345 total MIC measurements (**Fig. 2a Origin**). Some AMP/bacterium pairs occur multiple times due to overlap between databases and/or antimicrobial activity tests against multiple bacterial sub-strains [57].

We follow the data processing strategy in [57] to remove any peptides containing Cysteine for the exclusion of disulfide bonds. 35862 entries are obtained for further analysis (**Fig. 2a w/o_Cys**). We take the geometric mean when multiple measurements for an AMP/bacterium pair are present in the database similar as in [57], and finally get 25628 entries (**Fig. 2a w/o_Cys_Unique**). The preprocessed dataset consists of 3759 AMPs with their associated MIC values against *E. coli*, and 3373 peptides against *S. aureus*. These two bacteria carry the riches annotations in the database. Since *S. aureus* is one of the most prevalent pathogens threatening human life for their rapid increase in antimicrobial resistance [58] (**Fig. 2b**), this study uses *S. aureus* as an example to discuss the definition and prediction of AMPCliffs. More than 80% of the entries in the dataset have their lengths ranged from 7 to 25 amino acids (see **Fig. 2c**). So this study zooms into peptides whose length ranged within [7, 25] with their MIC values against *S. aureus*. The consolidated benchmark dataset consists of 2758 AMPs in total.

**Landscape of the benchmark dataset.** In order to figure out the challenges in the prediction of MIC values, we take the last layer of the pre-trained 33-layer ESM2 model as the representation of peptides, and employ t-SNE to visualize the two-dimensional distributions of the AMPs. Then we visualize the negative log10 of the MIC values as the z-axis of the overall landscape of the benchmark dataset. The unsmooth landscape intuitively illustrates the existences of many AC pairs (**Fig. 2e**). **Fig. 2f** shows 5 pairs of potential ACs and the complete list of the defined AMPCliff pairs is available in the Supplementary Table S6.

## Study rationale and experiment design

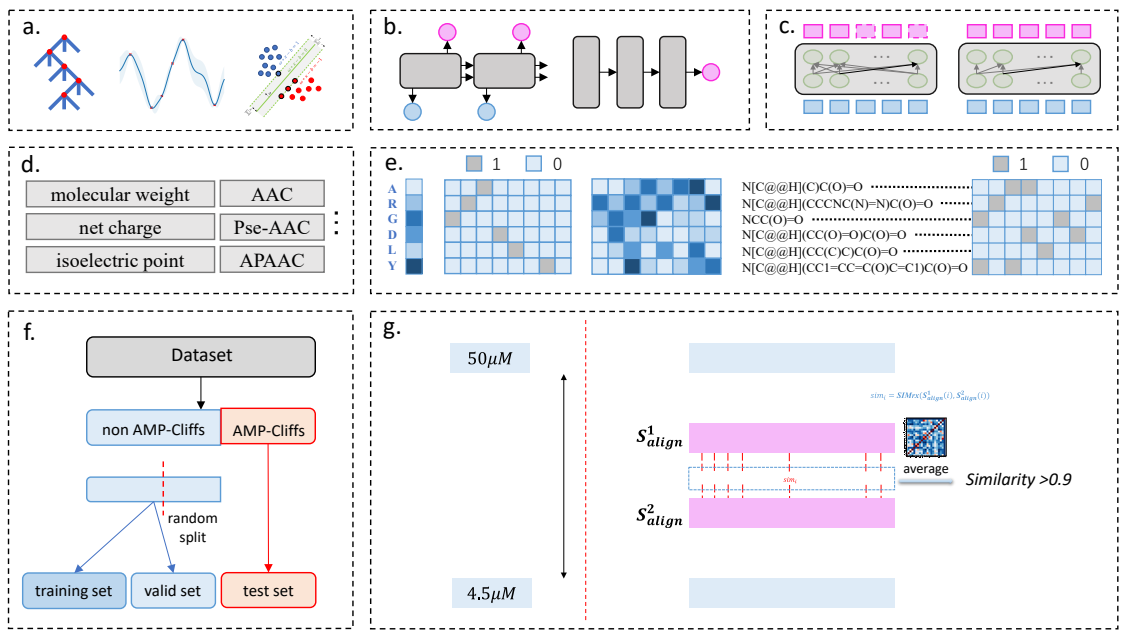

**Fig. 3 An overview of the quantitative definition and benchmarking of AMPCliff. a-c** are the model architectures used in this paper, including **a** machine learning methods: RF, XGBoost, GB, GP and SVM. **b** deep learning methods LSTM and CNN. **c** pre-trained language models like transformer encoder-based models BERT, ESM2 and transformer decoder-based model GPT2, ProGen2. **d-e** list the features used in this paper, including **d** fixed type representations, **e** dictionary index, one-hot encoding, word embedding and fingerprint. **f** the procedure of AC split. **g** an illustrative definition of AMPCliff.

**How to define activity cliff of AMPs?** Here we take two AMPs ALWKTLLKKVLKA as *sequence1* and ALWKTLLKKVLKAAA as *sequence2* to illustrate the procedure (**Fig. 3**). AMPCliff firstly aligns the two peptides using the Smith-Waterman algorithm. A substitution score matrix is used to calculate the similarity score between the two aligned peptides, and the residue-wise geometric mean is calculated as the final similarity score between these two peptides. **Fig. 3g** shows the MIC difference of *sequece1* and *sequence2* is around 11-fold, large enough to be treated as a pair of AMPCliff.

This study defines a pair of AMPs as AMPCliff, if these two AMPs with canonical amino acids are significantly similar to each other while their antimicrobial activity measurements, i.e., MIC values, have a large fold change (**Fig. 3g**). This is different to the definition of MMP-Cliff in [3] that the similarity between two molecules is measured on the atom-level and the potency change has to be at least 2 orders of magnitudes. We believe that the inter-molecular similarity in AMPCliff aligns more closely with the modular structure of amino acids in AMPs. The following sections will provide a detailed definition of AMPCliff.

**Does scaling law still work on AC property prediction?** The scaling law refers to that the performance of a language model (LM) improve consistently with the continually increased scale of this LM. This pattern was also observed in protein language models (PLMs) [41,44]. Therefore, this naturally raises a question: Does the scaling law of LMs and PLMs still hold effective in predicting the AC property in the AMPs with canonical amino acids? This question motivates us to evaluate the performance of transformer encoder- and transformer decoder-based models in increasing model sizes (illustrated in **Fig. 3c**). Moreover, some researchers have found that traditional machine learning methods outperformed advanced deep learning methods on AC property prediction tasks [1]. So we choose tree-based models, GP, and SVM as baselines (**Fig. 3a**). We also evaluate the effectiveness of published regression models for peptide design, including Huang et al.'s work [19], Pandi et al.'s work[20], and Schissel et al.'s work[40] (**Fig. 3b**). The details of the experimental settings of this work can be found in **Table 1**. If the corresponding methods were derived from the literature, the column "citation" will list its abbreviation in this paper and give the original reference. The hyperparameters of each model are listed in Supplementary Table S1.

**Table 1. The features and models evaluated in this paper.** If the corresponding methods were derived from literature, the column "citation" will list its abbreviation in this paper and give the original literature. Otherwise, denote "this paper". There are three categories of studies: machine learning (ML), deep learning (DL), and language model (LM).

| Category | Representation | models | citation |
| --- | --- | --- | --- |
| **ML** | Fixed type | support vector machine (SVM), linear regression (LR), L1, L2, ElasticNet, random forest (RF), gradient boost regressor (GB), | This paper |

|  |  | XGBoost, gaussian process regressor (GP) |  |
|---|---|---|---|
| **DL** | Dictionary index | LSTM | AMPSpace [19] |
| **DL** | One-hot encoding | LSTM | CellFree-rnn [20] |
| **DL** | One-hot encoding | CNN | CellFree-cnn [20] |
| **DL** | Fingerprint | CNN | peptimizer [40] |
| **LM** | Word embedding | BERT, ESM2 with 6,12,33 layers, GPT2, ProGen2 small, base and medium version. | This paper |

**How to evaluate the model performance on AC property prediction?** The dataset splitting strategy is essential to develop a stable model [17]. ACNet [11] utilized random split and target split to evaluate the model performance. But random split may separate a pair of AMPs one AMPCliff into training and test sets. This separation will reduce the influence of the ACs on the model performance and overestimate the capacity of the model on AC prediction tasks. While target split may introduce an out-of-distribution (OOD) issue into the model and amplifies the influence of ACs on the model performance.

This study introduces AC split, a data split strategy to ensure that AMPCliff pairs appear exclusively in the test set, and evaluates whether a peptide representation can effectively learn the high-frequency patterns commonly found in AMPCliffs from the low-frequency patterns in training set. The AC split strategy (**Fig. 3f**) firstly extracts all AMPCliff pairs with k-fold changes in the MIC values from the given dataset to form the test set. The remaining peptides are then divided into training and validation sets using a stratified random split strategy, consistent with the approach described in the literature [59].

# Results and Discussion

## AMPCliffs by Conserved Mutations

**Table 2 Ten example pairs of AMPCliffs with conserved mutations.** The two AMPs were aligned by the Smith-Waterman alignment with BLOSUM62 substitution matrix and set penalties of gap open/extension -11 and -1, which were default parameters in BLASTp [60]. The "Mutation" column indicates "(position in the alignment, amino acid in AMP1, amino acid in AMP2)". The detailed information can be found in the Supplementary Table S6.

| AMP1 | MIC1 | AMP2 | MIC2 | Mutation |
| --- | --- | --- | --- | --- |
| FLPIIGKLLSGLL | 4.45E-06 | FLPIVGKLLSGLL | 5.9E-05 | (5, I, V) |
| GLLKKIKWLL | 2.84E-05 | GLLKRIKWLL | 3.23E-06 | (5, K, R) |
| GWLDVAKKIGKAAFNVAKNFI | 2.23E-05 | GWLDVAKKIGKAAFNVAKNFL | 1.7E-06 | (21, I, L) |
| ILPWKWPWWKWRR | 2.58E-06 | LLPWKWPWWKWRR | 1.61E-05 | (1, I, L) |
| INLKAIAALAKKLL | 2.15E-05 | INLKAIAAMAKKLL | 1E-06 | (9, L, M) |
| KKKWLWLW | 6.74E-05 | KRKWLWLW | 8.23E-06 | (2, K, R) |
| KQKWLWLW | 3.79E-05 | KQRWLWLW | 3.7E-06 | (3, K, R) |
| RRLFRRILRWL | 5.5E-07 | RRLFRRILRYL | 4.5E-06 | (10, W, Y) |
| RRWWRWWR | 1.8E-06 | RRWYRWWR | 4.2E-05 | (4, W, Y) |
| RWWRWWR | 1.4E-05 | RWWRYWR | 2.07E-06 | (5, W, Y) |

We defined a pair of amino acids with a positive BLOSUM62 substitution score as the conserved substitution, and found out that even conserved substitutions can cause AC in AMPs. We illustrated 10 such AMP pairs as examples in Table 1. The complete list of AMPCliffs were released in Supplementary Table S6. For example, the pair of AMPs FLPIIGKLLSGLL and FLPIVGKLLSGLL only had a substitution from I to V (BLOSUM62 score 3) on the 5$^{th}$ residue, while their MIC values were increased from 4.45 µM to 59 µM, with ~~approximately~~ 13.2681-fold changes in the antimicrobial activities. Another pair of AMPs RRWWRWWR (MIC=1.80 µM) and RRWYRWWR (MIC=42 µM) carried a conserved substitution from W to Y and their antimicrobial activities received a 23.3116-fold change. The observations further supported the necessity of developing a precise peptide representation to reflect the subtle changes in AMPCliff.

## Measurement of similarity between two AMPs

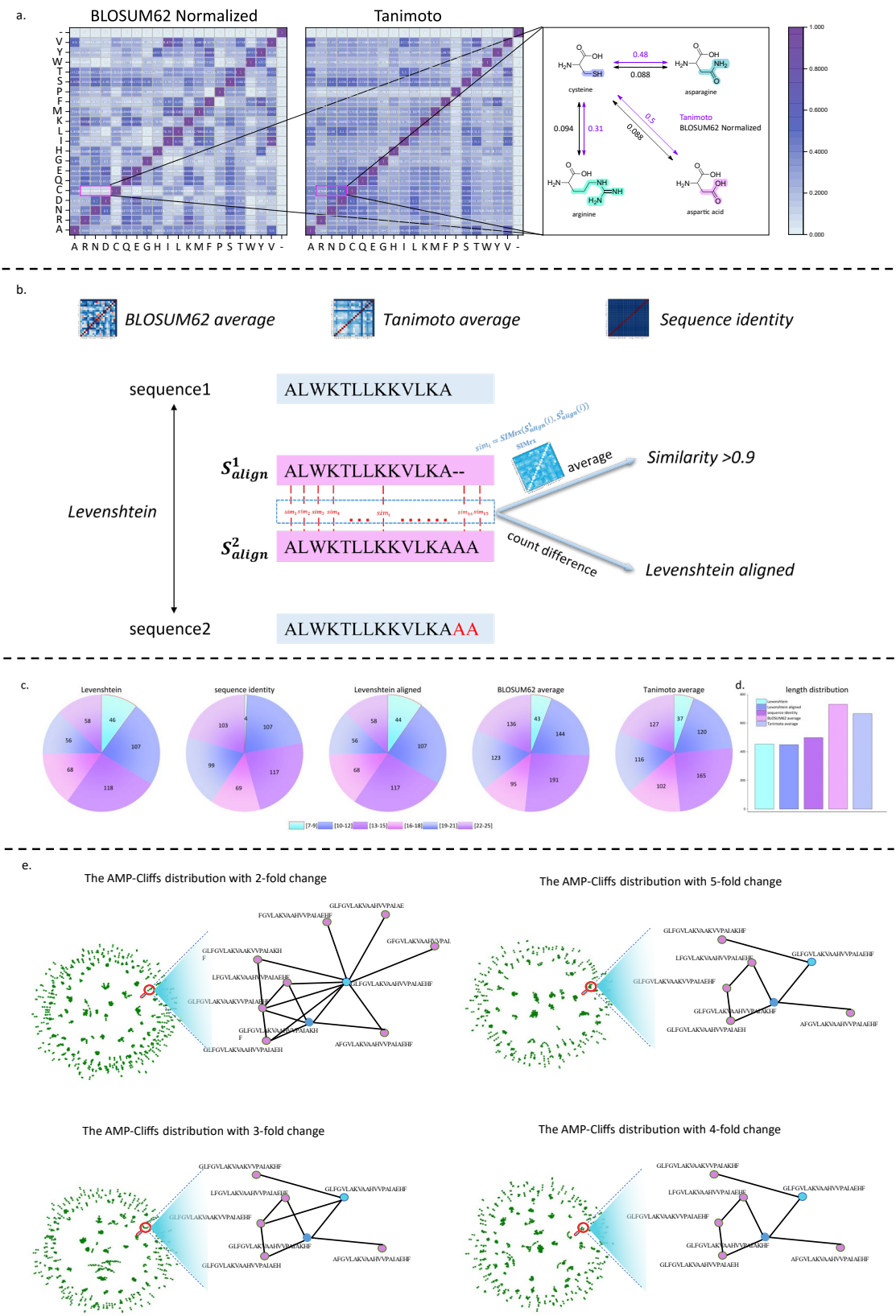

**Fig. 4 Similarity measurements between AMPs. a** The difference of amino-acid level similarity matrix measured by Tanimoto similarity and BLOSUM62 after normalization.

**b** A toy example of 5 similarity measurements, i.e., Levenshtein, Levenshtein_aligned, sequence_identity, Tanimoto_average, and BLOSUM62_average. **c** The numbers of AMPCliffs with 5-fold change in MIC values measured by Levenshtein, Levenshtein aligned, sequence identity, BLOSUM62 average, and Tanimoto average. Here we set 1 as the threshold for "Levenshtein" and "Levenshtein aligned", and 0.9 as the thresholds for the other 3 methods. **d** the total number of 5-fold change AMPCliffs under 5 similarity measurements. **e** Showcases of the variation of AMPCliffs as the fold change increased. Nodes represented sequences. If two nodes were marked as an AMPCliff pair, then they got an edge. The red node and the blue node represented the top 2 maximum degree nodes as fold change increased.

**Limitation of the MMP-Cliff definition in AMPs with canonical amino acids.** During our experimental investigations, we identified a limitation in the current definition of MMP-Cliff. This definition does not account for the structural similarity of substructures that transform within a pair of molecules, nor does it consider the evolutionary context of amino acids. Typically, substructures with structural similarities are likely to exhibit analogous functional characteristics. Therefore, a revised definition of activity cliffs (ACs) specific to AMPs with canonical amino acids is necessary.

**How to measure the similarity between two AMPs beyond MMP-Cliff?** This study investigated five commonly used similarity measurements between two peptides or two molecules to evaluate their performance, including Levenshtein distance (Levenshtein), Levenshtein distance between aligned sequences (Levenshtein_aligned), sequence_identity, an average of Tanimoto similarity between aligned sequences (Tanimoto_average), and an average of BLOSUM62 after normalized between aligned sequences (BLOSUM62_average). Two peptide sequences are aligned by the Smith-Waterman algorithm using biopython version 1.79 with Python version 3.8.18.

For a pair of sequences *sequence1, sequence2*, we denoted $S_1, S_2$ for short. Then we using sequence alignment algorithm to get the aligned sequence pair, denoted as $S_{align}^1$ and $S_{align}^2$ with the same aligned sequence length $L_a$. Furthermore, we used a similarity matrix $\boldsymbol{SIMrx} \in \boldsymbol{R}^{N \times N}$ to measure the similarity between two amino acids. For any two arbitrary amino acids *p* and *q*, we have their similarity value

$SIMrx(p,q) \in [0,1]$, and then we got the similarity between the two sequences $\frac{1}{L_a}\sum_{i=1}^{L_a} SIMrx(S_{align}^1(i), S_{align}^2(i))$.

**Is sequence identity an appropriate measure for peptide similarity in AMPCliff?**
Sequence identity is frequently used in protein design to demonstrate the novelty of discovered proteins [41,61], and can be measured using MMseqs2 [62], a widely-used clustering method for large protein datasets such as UniProt [63]. However, upon reviewing the UniProt database [64], we found that sequences shorter than 11 residues are excluded from the UniRef90 and UniRef50 datasets, which are clustered by MMseqs2 in the UniProt database. This poses a significant issue for rational peptide drug design, where peptides of 10-13 residues are common. Since many computational algorithms rely on data partitioning by MMseqs2, it is crucial to understand why MMseqs2 underperforms on short peptides and to design a new clustering method tailored to these short peptides.

Some researchers have already noted MMseqs2's limitations with short peptides [65,66] and have proposed alternative clustering algorithms, such as GraphPart [66] and GibbsCluster [67], to address the short peptide partitioning problem. GraphPart suggests that MMseqs2 selects representative sequences from clustering results and its maximum similarity criterion is not well-suited for homology partitioning [66]. Additionally, Hauser et al., has shown that the most sensitive tool for clustering short peptides of around 50 residues [65] is SWIPE [68], which is based on Smith-Waterman algorithm. GibbsCluster, on the other hand, is specifically designed to deconvolute multiple specificities in major histocompatibility complex (MHC) class I peptidome datasets [67].

To clarify this issue, we conducted a detailed analysis of MMseqs2's mathematical calculations. We identified two main drawbacks when applied to short peptides. First, MMseqs2's default module calculates k-mers before sequence alignment, a trade-off strategy that sacrifices precision for speed. While this is reasonable for long protein sequences, it is unnecessary for short peptides, which do not require such a trade-off. Second, MMseqs2 calculates sequence identity in two different ways [69]: (1) When using the alignment mode 3, it computes the number of identical aligned residues divided by the number of aligned columns, including columns containing gaps in either sequence. (2) Otherwise, it uses the local alignment bit score divided by the maximum length of the two aligned sequence segments. Neither method is ideal for measuring the similarity of short peptides. The first method is highly sensitive to sequence length, while the second method produces scores outside the [0, 1] range, making it challenging to set thresholds.

This study tries to combine the strengths of both methods by calculating the average similarity of each column in the two aligned peptide sequences. Then we need to score the similarity between two canonical amino acids in each column. A common method is to use the substitution scoring matrix BLOSUM62, though it uses integer values ranging from -4 to 11. Another method is the Tanimoto similarity, widely used in small-molecule drug design. The challenge is determining which method is more suitable for short AMPs with canonical amino acids. Additionally, as sequence length decreases, the precision of sequence alignment becomes a critical question.

**BLOSUM62 versus Tanimoto similarity on AMPCliff.** To compare the BLOSUM62 substitution matrix (with integer values ranging from -4 to 11) and the Tanimoto similarity of each amino acid pair based on ECFP (with float values ranging from 0 to 1) on the same scale, we employ a normalization strategy akin to the MaxMinScaler in the Python sklearn package, with an additional step to ensure the symmetry of the formula:

$$M'_{\text{new}} = \frac{M - \min(M)}{\max(M) - \min(M)} \tag{1.}$$

$$M_{\text{new}} = \frac{M'_{\text{new}} + {M'_{\text{new}}}^{\text{T}}}{2} \tag{2.}$$

where M represents the original scoring matrix, and $M_{\text{new}}$ is the normalized version. **Fig. 4a** illustrates the details of the normalized BLOSUM62 matrix (referred to as "BLOSUM62 average") and Tanimoto similarity matrix. We observe that certain amino acids exhibit similar patterns across both matrices. For example, the normalized BLOSUM62 values for phenylalanine (F) with tryptophan (W) and tyrosine (Y) are 0.42 and 0.65, respectively, with phenylalanine (F) being most similar to tyrosine (Y) compared to other amino acids. This similarity is mirrored in the Tanimoto similarity matrix, where the similarity values for phenylalanine (F) with tryptophan (W) and tyrosine (Y) are 0.5 and 0.7, respectively, again showing phenylalanine (F) as most similar to tyrosine (Y).

However, significant differences also emerged. For instance, the normalized BLOSUM62 values between cysteine (C) and arginine (R), asparagine (N), and aspartic acid (D) are all below 0.1. In contrast, the Tanimoto similarities between cysteine (C) and arginine (R), asparagine (N), and aspartic acid (D) are 0.3, 0.5, and 0.52, respectively. This discrepancy arises because Tanimoto similarity does not account for structural similarity at the atomic level, and assumes that all atomic substitutions have the same impact.

As shown in **Fig. 4a**, cysteine (C) contains sulfur (S) and often forms disulfide bonds, while arginine (R), asparagine (N), and aspartic acid (D) have markedly different side chains and chemical properties. Arginine (R) is basic with an amino group (-NH2), asparagine (N) is amidic with an amide group (-CONH2), and aspartic acid (D) is acidic with a carboxyl group (-COOH). These fundamental differences lead to distinct physicochemical properties that Tanimoto similarity fails to capture. The MMP-Cliff definition suffers from a similar issue as Tanimoto similarity, i.e., treating atom-level differences uniformly.

In summary, we combine the advantages of the two sequence identity calculation methods and propose the "BLOSUM62 average" as a more suitable similarity measure for short-length peptides with canonical amino acids.

**Global alignment versus local alignment.** Both global and local alignments are widely used for sequence alignment. Global alignment compares two sequences across their entire lengths, and is suitable for sequences expected to share similarity throughout their entirety [70]. Global alignment maximizes regions of similarity and minimizes gaps using scoring matrices and gap penalties provided to the algorithm. In contrast, local alignment identifies regions of local similarity and does not require aligning the entire length of the sequences. Peptides are much shorter than protein sequences, and typically consist of fewer than 50 amino acids. In this study, we analyze sequences ranging from 7 to 25 amino acids in length, and it's necessary for a reconsideration of which type of alignment is most appropriate.

Teufel et al. proposed GraphPart [66] to utilize the global Needleman-Wunsch alignment through the EMBOSS module "needle" with default parameters. However, Baylon et al. [71] systematically evaluated the performance of global and local alignments on therapeutic peptides in their development of PepSeA, and concluded that local alignment was more accurate for these shorter sequences.

We observe that GraphPart truncated protein datasets, such as SignalP 5.0 [72] (to obtain signal peptides with 70 N-terminal amino acids) and NetGPI [73] (to obtain GPI anchors with 100 C-terminal amino acids). These truncated peptides are longer and have fixed lengths compared to the sequences used in this study (7 to 25 amino acids). The sequence length distribution in PepSeA [71], which ranged from 9 to 18 amino acids, is much closer to our research scope. Therefore, we adopted the local alignment approach from PepSeA.

In summary, this study investigates AMPCliffs in peptides with canonical amino acids, and employs the Smith-Waterman local alignment algorithm with the BLOSUM62 substitution matrix. We set the gap opening and gap extension penalties to -11 and -1, which are the default parameters in BLASTp [60].

## Measurement of bioactivity changes between two AMPs

**Setting the fold-change threshold for bioactivities.** The definition of an MMP-Cliff [3] requires that the potency difference between two compounds meeting the structural criteria must be at least two orders of magnitude. However, the measurement of bioactivities in AMPs uses a different metric, minimum inhibitory concentration (MIC), to assess antimicrobial potency. A lower MIC value indicates higher antimicrobial activities.

Mouton et al. [74] pointed out that the ISO 20776-2 standard for MIC reproducibility allows for an acceptable deviation of one dilution from the mode in 95% of cases, corresponding to a range of at least two 2-fold dilutions. To determine an appropriate fold-change threshold for AMPCliffs, we evaluated how different minimum MIC differences in AMPCliff pairs, ranging from a 5-fold to a higher fold change, affect model performance using the Recall metric (see "Performance evaluation of AMPCliff predictions" for details). For the definition of Recall, please see **Evaluation Metrics**.

An MIC value with the unit mole (M) is firstly converted to –log(MIC) (the logarithm base 10). We defined the MIC values of any two sequences as $MIC_1$ and $MIC_2$, with $MIC_1 \geq MIC_2$. The fold change is defined as:

$$\frac{\text{MIC}_1}{\text{MIC}_2} \geq \tau \qquad (3.)$$

$$-\log\left(\frac{\text{MIC}_1}{\text{MIC}_2}\right) = -\log\text{MIC}_1 - (-\log\text{MIC}_2) \leq -\log(\tau) \qquad (4.)$$

$$|-\log(\text{MIC}_1) - (-\log(\text{MIC}_2))| \geq \log(\tau) \qquad (5.)$$

In order to define the fold change threshold properly, we set τ = 2, 3, 4, and 5 to establish the dataset for model training and performance comparison with recent state-of-the-art methods for the AMPCliff prediction task. (See **Performance evaluation of AMPCliff predictions**) Note that a larger -log(MIC[M]) value indicates better antimicrobial activities.

**Experimental data supporting theoretical analysis.** To ensure our methods are applicable in real-world scenarios, we compared the number of detected 5-fold change AMPCliffs across different sequence length regions using various similarity metrics: Levenshtein, Levenshtein_aligned, sequence_identity, BLOSUM62_average, and Tanimoto_average. **Fig. 4b** provides a schematic of these measurement procedures, while **Fig. 4c-d** present the numbers of detected 5-fold AMPCliffs across different length ranges.

Levenshtein distance is also known as edit distance, and it measures the difference between two sequences by counting the minimum number of single-character edits (insertions, deletions, or substitutions) required to convert one sequence into the other. The difference between "Levenshtein" and "Levenshtein_aligned" lies in whether the sequences are aligned prior to comparison (see **Fig. 4b**). Our results show that these two metrics yield similar outcomes (see **Fig. 4c**).

The distinction between "Levenshtein_aligned" and "sequence_identity" is that sequence_identity averages similarity over the entire sequence length (see **Fig. 4b**). This averaging operation biases sequence identity toward identifying longer sequences. For example, if two aligned sequences $S_{\text{align}}^1$ and $S_{\text{align}}^1$ differ by only one residue, the sequence identity will be only 0.9 if the aligned sequences are 10 residues. This bias explains why AMPCliffs identified by sequence identity tend to be longer and why few sequences in the [7-9] length range are detected (see **Fig. 4c**). This also suggests that MMseqs2 is not suitable for clustering short peptides, a conclusion that aligns with experimental findings in the literature [64,65].

The length distributions of AMPCliffs identified by "BLOSUM62_average" and "Tanimoto_average" are similar to each other, with "BLOSUM62_average" generally detecting a high count (see **Fig. 4d**). Each metric has its strengths and limitations. BLOSUM62_average incorporates the prior knowledge of medicinal chemists and evolutionary information, but cannot be extended to the peptides with non-canonica amino acids. Tanimoto_average treats all atomic substitutions equally and disregards evolutionary information. But it can be easily applied to peptides with non-canonical amino acids.

Therefore, this study presents two versions of the AMPCliff identification models. For tasks involving canonical amino acids, the BLOSUM62 version is recommended. For tasks involving non-canonical amino acids, the Tanimoto version is preferred if a suitable sequence alignment algorithm for non-canonical peptides is available. To our knowledge, PepSeA [71] is the only sequence alignment algorithm currently available for such peptides. This paper primarily reports results from AMPCliff predictions using the "BLOSUM62_average" metric, with the detailed results for both "BLOSUM62_average" and "Tanimoto_average" metrics provided in Supplementary Tables S2-S4 and S7-S8.

In summary, our experimental results confirm that sequence identity is not a suitable metric measuring similarity between short AMPs with canonical amino acids. Instead, our proposed "BLOSUM62_average" measurement can detect the highest number of AMPCliffs.

## Quantitative definition of AC in AMPs

We further propose a new quantitative definition of AC in AMPs with canonical amino acids, termed AMPCliff. The principles of AMPCliff align with common understanding of AC that structurally similar molecules are active against the same bacteria but exhibit significantly different MIC values. We set a 2-fold change as the minimum threshold for MIC differences. The similarity between two AMPs is defined as follows:

Given a pair of peptide sequences $S_1$ and $S_2$, we first use the Smith-Waterman alignment algorithm to obtain the aligned sequences, denoted as $S^1_{align}$ and $S^2_{align}$, of the same length, $L_a$. We utilize a similarity matrix $\mathbf{SIMrx} \in \mathbf{R}^{N \times N}$ to measure the similarity between two residues in the alignment. For a pair of matched residues $p$ and $q$ in the alignment, their similarity value is defined as $\mathbf{SIMrx}(p,q) \in [0,1]$. The overall similarity between the two aligned peptide sequences is then calculated as:

$$\frac{1}{L_a} \sum_{i=1}^{L_a} \mathbf{SIMrx}(S^1_{align}(i), S^2_{align}(i))$$

**Fig. 3g** illustrates the AMPCliff definition procedure using the two AMP sequences ALWKTLLKKVLKA and ALWKTLLKKVLKAAA as examples.

The key difference between AMPCliff and MMP-Cliff lies in the similarity measurement. AMPCliff first employs sequence alignment to align the two molecules, and then uses a similarity matrix to calculate the similarity of residues at each aligned position. The final similarity is obtained by taking the geometric mean of these values.

Having established the definition of AC in AMPs with canonical amino acids (AMPCliff), the next question is how can we evaluate model performance in predicting AMPCliffs?

**Performance evaluation of AMPCliff predictions**

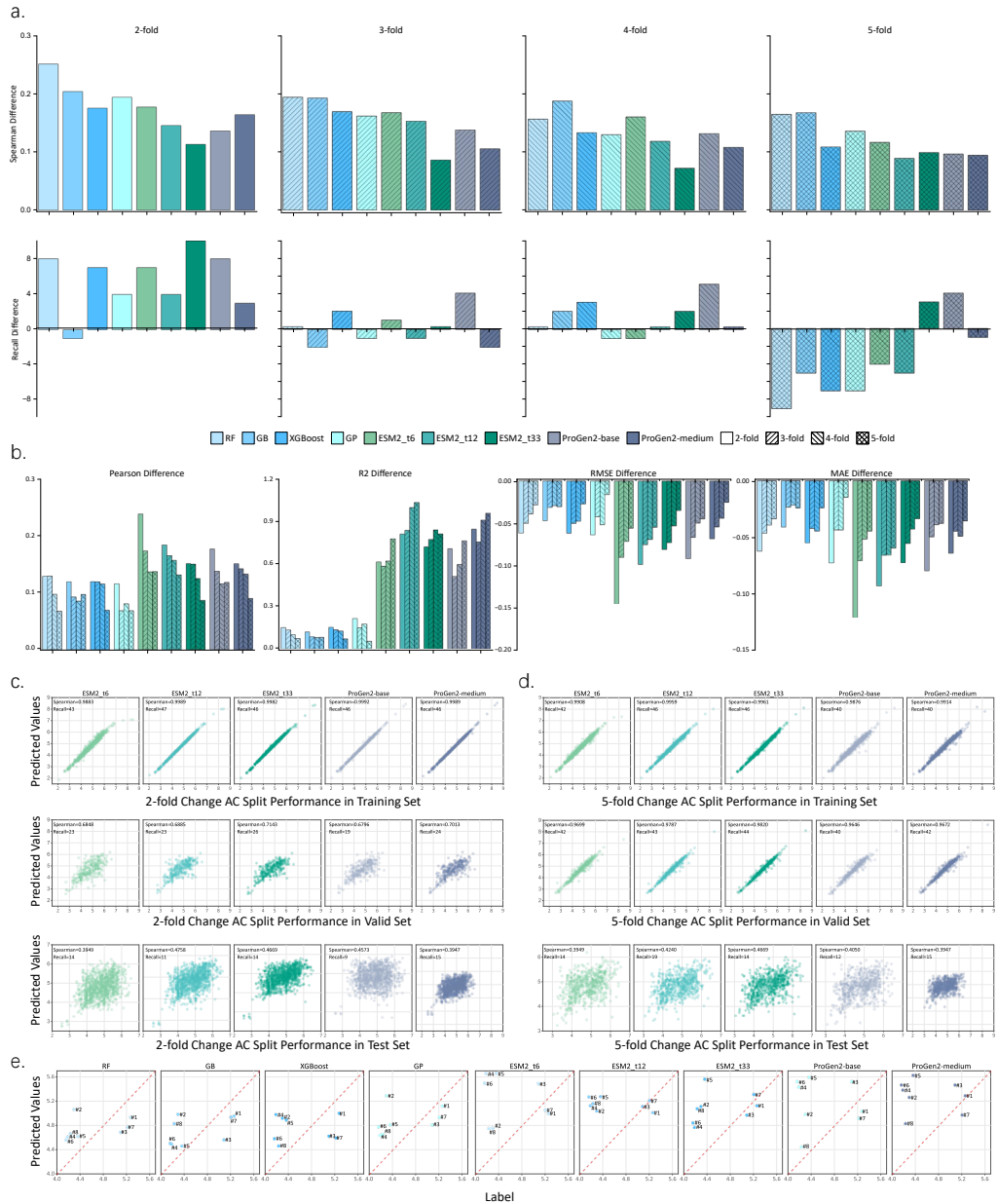

**Fig. 5** Performance comparison between random split and AC split methods. Panels (**a-b**) show the difference in Spearman, Recall (**a**) and other performance metrics (**b**) Metric(random split) – Metric(AC split), varying from 2-fold to 5-fold changes in the MIC threshold. Panels (**c-d**) present scatter plots for the training/validation/test sets under 2-fold (**c**) and 5-fold (**d**) in the MIC threshold and AC splits. Panel (**e**) displays a scatter plot of ACs with 5-fold changes in the MIC threshold from **Fig. 2**, where the red line represents the diagonal line *y=x*. The label is the -log(MIC[M]) value of each AMP.

We compare the performance of a 4:1 stratified random split (referred to as "random split") with our proposed AC split, varying the MIC threshold from 2-fold to 5-fold changes. We denote the AC split strategy with the *k*-fold changes in the MIC threshold as *k*-fold AC split, where *k*=2, 3, 4, or 5. The model does not encounter high-frequency data in the AMPCliffs in the training set when using the AC split strategy, while some sequences in the AMPCliffs may appear in the training set of the random split strategy. The performance of the random split strategy is expected to be overestimated. However, our observations suggest that this common assumption may not always hold true in practice.

As shown in **Fig. 5(a-b)**, the choice of performance metric can significantly influence the perceived model performance. For instance, the difference in the metric SCC (upper panel of **Fig. 5a**) and other metrics like PCC and R2 (**Fig. 5b**) between random split and AC split with 2/3/4/5-fold changes are generally positive (negative for RMSE and MAE). This indicates that random split is indeed overestimated, consistent with the literature[17]. However, the Recall metric (lower panel of **Fig. 5a**) does not follow the same trend. As the fold change increases from 2 to 5, the model's performance with random split initially outperforms the AC split with 2-fold change, then becomes competitive to the AC splits with 3- and 4-fold changes, and eventually underperformed relative to the AC split with 5-fold change. This phenomenon is intriguing because the test set of the AC split with 2-fold change includes AMPCliffs with 2-fold and larger changes, whereas the test set of the AC split with 5-fold change only includes AMPCliffs with MIC changes of 5-fold or greater. This result suggests that while the random split can predict low-frequency (2-fold to 3-fold) and relatively higher frequency data (3-fold to 4-fold) well, it struggles with very high-frequency data (5-fold and above).

If we change the data split strategy from random split to AC split, the models may perform poorly in predicting exact MIC values (with lower Pearson, R2, RMSE, and MAE compared to random split) and general MIC ranking (lower Spearman correlation). However, they can improve the Recall of the top 50 sequences, which is crucial in practice as we often focus on the top *k* predictions. Consequently, we determined that a 5-fold change is the maximum threshold for defining an AMPCliff.

We further examine the results of the pre-trained language models (LMs) on the 2-fold and 5-fold AC splits across training, validation, and test sets (**Fig. 5c-d**). The validation set performance for the 2-fold AC split is less consistent compared to the 5-fold AC split, which is understandable given that a 2-fold change represents the minimum reproducibility tolerance for the same peptide on a MIC detection device [74]. This suggests that the training set for the 2-fold AC split contains more systematic noise. Conversely, the 5-fold AC split results indicate that the LMs are well-trained on peptides within the 5-fold change range. Although these models are poor at predicting ACs, they have still learned some general knowledge from non-AC peptides (low-frequency data) to AC peptides (high-frequency data), supporting the rationale for using a 5-fold threshold in our AMPCliff definition.

## Case studies of AMPCliff predictions

We zoom into specific samples to demonstrate how they are predicted with the 5-fold AC split strategy by four ML models and five LM models. **Fig. 5e** shows the dot plots of the labels (-log(MIC[M])) on the horizontal axis and their predicted values on the vertical axis. A stronger antimicrobial activity is supported by a smaller MIC value, and the corresponding -log(MIC[M]) value is larger than that of a weak AMP.

Three peptides with smaller MIC values (#1 with 5.5$\mu M$, #3 with 8.2$\mu M$, #7 with 6.2$\mu M$) in the AMPCliff pairs are commonly located at the right bottom area under the red line *y=x* by all the ML modes. This suggests that the ML models predict these three strong antimicrobial peptides as weak ones.

Finally, we zoom into specific samples to see how they perform with the 5-fold AC split by 4 ML methods and 5 LMs. As shown in **Fig. 5e**, it turns out that three peptides with lower MIC values in the AMPCliff pairs (here we use -log(MIC[*M*]) as labels, which means the higher label values in the figure) are #1 with 5.5$\mu M$, #3 with 8.2$\mu M$, #7 with 6.2$\mu M$ and commonly locate at the right bottom of the *y=x* red line by all ML methods. Besides, except for XGBoost, all the other three ML methods have correctly predicted the relative order of these 3 peptides well.

In contrast, none of the LM models have accurately predicted their relative orders. Besides, for the higher MIC values in the AMPCliff pairs, #2 with 50.0$\mu M$, #4 with 64.4$\mu M$, #5 with 68.4$\mu M$, #6 with 43.0$\mu M$, #8 with 58.0$\mu M$, their pairwise order is hard to predict precisely for all models. On the one hand, this suggests that the data labels with systematic noise are not suitable for using metrics like RMSE, MAE, Pearson, and R2 to evaluate model performance in practice. Instead, a coarse-grained evaluation method is needed, like the Recall metrics for local evaluation and Spearman correlation coefficients for global evaluation. On the other hand, recent LMs cannot predict the order of ACs more accurately than MLs, revealing one of the drawbacks of LMs. The next section systematically compares the performance of LMs with MLs.

As we discussed in the former section, the choice of performance metrics could influence the perceived model performance. SCC is much easier to be affected by data split strategy than Recall. For ML methods, the Recall of AC split with 5-fold change even gets higher than random split (see **Fig. 5a-b**), which guides us to zoom into specific samples to demonstrate how they are predicted with the AC split strategy with 5-fold change by four ML models. We also plot the results of five LM models (three MLMs: ESM2 with 6 layers, 12 layers and 33 layers. two GLMs: ProGen2 with base and medium version) for comparison. **Fig. 5e** shows the dot plots of the labels (-log(MIC[M]) values) on the horizontal axis and their predicted values on the vertical axis. A stronger antimicrobial activity is supported by a smaller MIC value, and the corresponding label (-log(MIC[M]) value) is larger than that of a weaker AMP.

Firstly, we find out that although ML methods are bad at predicting AMPCliff pairs accurately, which is common in all existing algorithms, they are good at predicting relative orders by AC split. By showing the prediction results of seven cases which are consistent with **Fig. 2f**, we group them into strong MIC peptides and weak MIC peptides. Table 3 shows the details of these 7 peptides. All the four ML methods do not predict the label -log(MIC[M]) values accurately, and tend to predict larger values for weak AMPs (located at the left upper area under the red line *y=x*) and smaller values for strong AMPs (located at the right bottom area under the red line *y=x*) (see **Fig. 5e**). However, if we dig into the details of each group, it turns out that random split is more likely to fail to discriminate the strong and weak AMPs than AC split. For random split, the prediction values of the seven samples by all four ML methods (in the upper row of **Fig. 5e**) are mixed in the same region, failing to distinguish the strong AMP group from the weak group. Whereas for AC split, the two groups of AMPs are easy to be separated especially by RF and GP (in the lower row of **Fig. 5e**). The strong AMP group generally has larger predicted -log(MIC[M]) values than the weak AMP group except for the #2 sample.

Furthermore, we investigate the performance of LMs on the seven samples in Table 3. It turns out that LMs tend to generally predict the seven samples with larger -log(MIC[M]) values by AC split (in the lower row of **Fig. 5e**) than by random split (in the upper row of **Fig. 5e**). It is beneficial for the strong AMP group, but not for the weak AMP group. Besides, the order of the seven AMPs is collapsed by random split and AC split. This reveals one of the drawbacks of LM representation. The next section systematically compares the performance of LMs with MLs.

Table 3. The details of the 7 example AMPs.

| Group | ID | AMP | MIC ($\mu M$) | Labels(-log(MIC[M])) |
|---|---|---|---|---|
| Strong AMPs | #1 | AFHHIFRGIVHVGKTIHRLVTG | 5.5 | 5.26 |
| | #3 | AGRGKQGGKVRAKAKTRSS | 8.2 | 5.08 |
| | #7 | AGYLLGKINLKALAALAKKIL | 6.2 | 5.21 |
| Weak AMPs | #6 | GRGKQGGKVRAKAKTRSS | 43.0 | 4.37 |
| | #2 | FFHHAFRGIVHVGKTIHRLVTG | 50.0 | 4.30 |
| | #8 | AGYLLGKINLKPLAALAKKIL | 58.0 | 4.23 |
| | #4 | DGRGKQGGKVRAKAKTRSS | 64.4 | 4.19 |
| | #5 | KGRGKQGGKVRAKAKTRSS | 68.4 | 4.16 |

## Learned representations versus fixed type representations

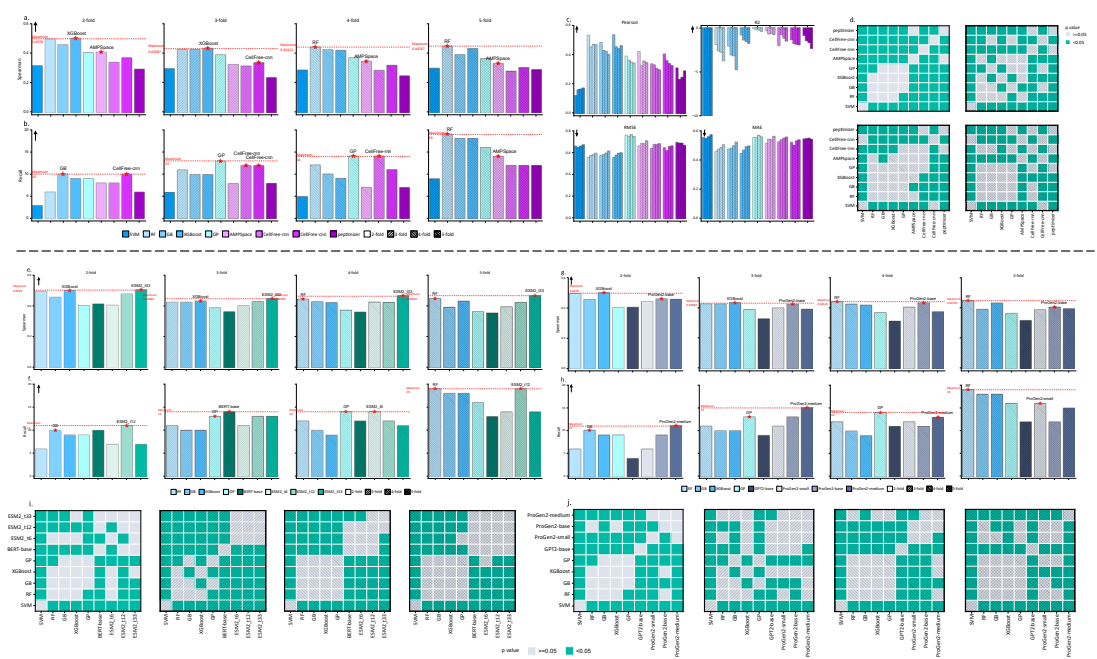

**Fig. 6 Prediction performance with the benchmark dataset by AC split.** The performance metrics (a) Spearman correlation coefficient and (b) recall of five ML methods SVM, RF, GB, XGBoost, and GP on fixed type representations and four DL methods AMPSpace, CellFree-rnn, CellFree-cnn and peptimizer on the AMPCliffs defined by "BLOSUM62 average" with 2-fold, 3-fold, 4-fold and 5-fold change. (c) Other prediction performance metrics of the models. (d) Statistically predicted value significance for pairwise model comparison by Mann-Whitney U test. (e) Spearman and (f) recall of five ML methods SVM, RF, GB, XGBoost, and GP on fixed representations and four masked language model methods BERT, ESM2_t6, ESM2_t12 and ESM2_t33 on the AMPCliffs with 2-fold, 3-fold, 4-fold and 5-fold change. (g) Spearman correlation coefficient and (h) recall of five ML methods SVM, RF, GB, XGBoost, and GP on fixed representations and four generative language model methods GPT, ProGen2-small, ProGen2-base and ProGen2-medium on the AMPCliffs with 2-fold, 3-fold, 4-fold and 5-fold change. Statistically predicted value significance for pairwise model comparison by Mann-Whitney U test on (i) the four masked language models (MLMs) BERT, ESM2_t6, ESM2_t12 and ESM2_t33 and (j) four generative language models (GLMs) GPT, ProGen2-small, ProGen2-base and ProGen2-medium.

The learned representations refer to those peptide features derived from the pre-trained DL models, while the fixed type representations are defined and calculated by manually formulated equations. To check if learned representations outperform fixed type representations, we compare ML models GP and SVM with published deep learning methods AMPSpace, CellFree-rnn, CellFree-cnn and peptimizer on AMPCliffs defined by "BLOSUM62 average" with 2-fold, 3-fold, 4-fold and 5-fold changes. Notice that we add a new model SVM into ML methods for comparison, for checking whether it is suitable for AMPCliff prediction.

Firstly, we consider Recall and SCC as the performance metrics, since we care more about the relative order of the predicted -log(MIC[M]) values of an AMPCliff pair than their specific prediction values in the process of drug discovery. Besides, we calculate Recall involved sorting both the top 50 of the label values and their predicted -log(MIC[M]) values in descending order, and a larger value means the better antimicrobial activity. As shown in **Fig. 6a-b,** RF gets the best SCC value with 4-fold and 5-fold change. XGBoost gets the best SCC value with 2-fold and 3-fold change. Meanwhile, RF and GB achieve the best Recall values with 5-fold dilution and 2-fold change, respectively. GP gets the best Recall value with 3-fold and 4-fold change. SVM performs poorest compared with the other ML methods, and none of the four DL methods have surpassed the best traditional methods, which is consistent to the literature[18].

Moreover, when we gradually add some lower frequency signals like no bigger than 3-fold and 4-fold AMPCliffs into the training set, while the SCC decreases, the Recall raises gradually. Although SCC doesn't follow the same trend as Recall, it is worthy to know that except for the 2-fold change, the others had a similar trend. The data indicates that models can learn some inherent patterns from the low-frequency signals to successfully predict the high-frequency signals, and the models can predict the top AMPs more effectively as seeing more high-frequency data. The details of the results for each model can be found in Supplementary Table S2.

Furthermore, we select the best 4 ML methods, RF, GB, XGBoost and GP for comparing prediction performance with 4 MLMs: BERT bae version, ESM2 with 6 layers, ESM2 with 12 layers and ESM2 with 33 layers. As shown in **Fig. 6e-f,** the largest model ESM2 with 33 layers beat the ML methods and reaches the best performance in SCC among all fold changes. At the meantime, in terms of the metric Recall, although MLMs especially ESM2 don't follow the same trend as they are in SCC, they get the better Recall compared with the ML methods in general. Another noteworthy result is the performance of BERT pretrained by English text. Although its SCC gets the worst among all of the MLMs, its Recall reaches competitive performance with ESM2. And its predicted values don't have a significant difference compared with ESM2 with 12 layers among all fold changes (**Fig. 6i**).

Meanwhile, we compare the four best ML methods with four GLMs: GPT2, Progen2 small version, base version and large version. As shown in **Fig. 6g-h,** it turns out that GLMs perform poorer than ML methods in SCC values and only the Recall of the medium size of ProGen2 beats ML methods on the 2-fold and 3-fold change. The second-best Recall on the 2-fold and 3-fold change are still the ML methods GB and GP, respectively.

Therefore, ESM2 with 33 layers beats fixed representations in SCC and Recall, whereas ProGen2 and DL methods remain to be further improved. We speculate that ESM2 captures some structural information from peptide sequences.

Firstly, fixed type representations with ML methods outperform the conventional DL methods. We continue to consider Recall and SCC as the performance metrics since the relative order of an AMPCliff pair's predicted values is more valuable than the values themselves in the drug discovery process. We calculate Recall for the top 50 of the label values and their predicted values (For the definition of Recall, please see **Evaluation Metrics**.). Since we use -log(MIC[M]) as the label value, a larger value means the better antimicrobial activity of the AMP. As shown in **Fig. 6a-b,** RF gets the best SCC value with 4-fold and 5-fold changes. XGBoost achieves the best SCC value with 2-fold and 3-fold changes. Meanwhile, RF and GB achieve the best Recall with 5-fold change and 2-fold change, respectively. GP gets the best Recall value with 3-fold and 4-fold changes. SVM performs poorest compared with the other ML methods, and none of the four DL methods have outperformed the best traditional methods (namely, the fixed type representations with ML methods.), which is consistent with the literature[18]. All of the best performances by AC split across 2/3/4/5-fold changes belong to the fixed type representations with ML methods.

Moreover, it turns out that the fixed type representations with ML methods can transfer some general patterns from low-frequency data to high-frequency data. Table 4 shows the Recall of the fixed type representations with each ML method by the AC Split with 2/3/4/5-fold changes, namely gradually adding more low-frequency data to the training data. Except for XGBoost, the Recall of all the other ML methods increases or stays at the same value as the training data see more low-frequency data. Besides, the XGBoost gets 9 for 4-fold change and 10 for 3-fold change, which is quite close to each other. More evidence may be found in the Supplementary Table S2.

Table 4. The metric Recall of the fixed type representations with each ML method by the AC Split with 2/3/4/5-fold changes. The columns are the AC Split with 2-fold to 5-fold change, and a row represents each ML method. The ML methods use the same fixed type representation features. For the definition of Recall, please see **Evaluation Metrics**. 【这里的 Recall 怎么全是证书？！请定义清楚！】

| Fold | 2 | 3 | 4 | 5 |
|---|---|---|---|---|
| SVM | 3 | 6 | 5 | 9 |
| LR | 9 | 9 | 10 | 12 |
| **L1** | 9 | 8 | 11 | 10 |
| **L2** | 8 | 8 | 8 | 15 |
| **ElasticNet** | 7 | 8 | 8 | 10 |
| **RF** | 6 | 11 | 12 | 19 |
| **GB** | 10 | 10 | 10 | 18 |
| **XGBoost** | 9 | 10 | 9 | 18 |
| **GP** | 9 | 13 | 14 | 16 |

The learned representations based on LMs, especially ESM2, have the potential to beat the fixed type representations. Since SVM is the poorest model, we remove it from further analysis. Then we compare the four ML methods, RF, GB, XGBoost, and GP with 4 MLMs: BERT bae version, ESM2 with 6 layers, ESM2 with 12 layers, and ESM2 with 33 layers. As shown in **Fig. 6e-f,** the largest model ESM2 with 33 layers beats the ML methods and achieves the best performances in SCC among all fold changes. Besides, there is an interesting ascending trend in the metric SCC of ESM2 as the model size gets larger. In the meantime, in terms of Recall, MLMs get the best Recall compared with the ML methods in general, but ESM2 don't have the same ascending trend as it does in SCC. Another noteworthy result is that the prediction values of the English text-based pre-trained model BERT perform similarly to ESM2 with 12 layers. Although its SCC is the worst among all the MLMs, its Recall metric is competitive to ESM2. **Fig. 6i** calculates the statistical significance of the null hypothesis of the predicted values of two models are the same by the Mann-Whitney U test. The p-value smaller than 0.05 means that the predicted values of the compared models is significantly different. It turns out that the predicted values of BERT with the base version don't have a significant difference compared with ESM2 with 12 layers among all fold changes.

Meanwhile, we compare the four ML methods with four GLMs: GPT2, Progen2 small version, base version, and large version. As shown in **Fig. 6g-h,** it turns out that GLMs perform worse than ML methods in SCC and only the Recall of the medium size of ProGen2 beats ML methods on the 2-fold and 3-fold changes, and the second-best Recall on the 2-fold and 3-fold change are still the ML methods GB and GP.

ESM2 with 33 layers beats the fixed types representations in SCC and the learned representations based on MLMs can beat the fixed type representations in Recall, while ProGen2 or the conventional DL methods remain to be further improved. We speculate that the MLM-based learned representations can be good solutions for solving the AMPCliff prediction task. This training paradigm can capture some structural information from peptide sequences, as claimed in ESM2[41] and ESM3[75].

## Does Scaling Law Still Work?

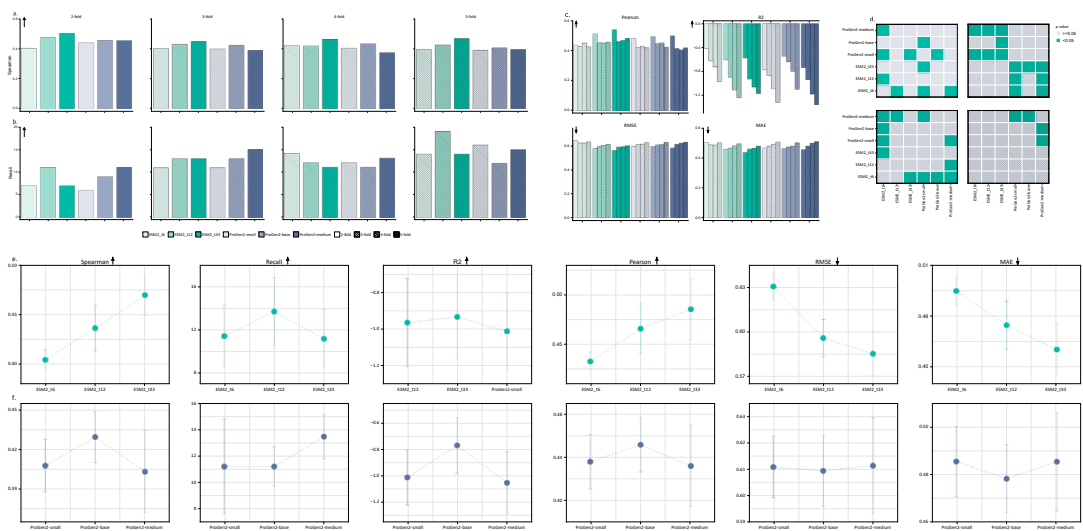

**Fig. 7 Evaluating the prediction performance of MLMs and GLMs.** (a) SCC and (b) Recall of three MLMs and three GLMs on the AMPCliffs defined by "BLOSUM62 average" with 2-5 fold changes. **(c)** Other prediction performance metrics of the models. (d) Statistical predicted value significance for pairwise model comparison by Mann-Whitney U test. **T**he average performance of each (e) MLM or (f) GLM with 2-5 fold changes.

Here we take the scaling laws defined by OpenAI into consideration[76]. Model performance heavily depends on scale, which consists of three factors: the number of model parameters (excluding embeddings), the size of the dataset, and the amount of computing power used for training[76]. Since we utilize the pre-trained language models to fine-tune the AMPCliff prediction task, and evaluate whether it has an ascending trend on each performance metric as the model scale gets larger.

To study whether the scaling laws still work on the AMPCliff prediction task, we evaluate the prediction performance of ESM2 with 6/12/33 layers and ProGen2 small/base/medium versions under the "BLOSUM62 average" condition by AC split. The analysis aims to answer the question: does scaling law still work on the AMPCliff prediction task? Note that BERT and GPT are removed from this analysis because they are not in the ESM2 or ProGen2 series.

Firstly, we investigate their performance in the metrics SCC and Recall, and find out that ESM2 has an ascending trend on predicting the relative order of AMPCliffs. As shown in **Fig. 7a**, it has an obvious SCC improvement as the model size of ESM2 increases. This trend exists almost in all fold changes. Whereas it disappears in the Recall metric (see **Fig. 7b**), which suggests that the top 50 samples in the test dataset don't perform better as the model size of ESM2 increases. At the same time, GLMs don't show a similar trend as MLMs, i.e., there is neither significant increasing trend in SCC nor Recall except for the 2- and 3-fold changes. This suggests that GLMs are possibly good at predicting the top 50 samples in AMPCliffs other than the general performance. Besides, if we compare the other performance metrics as shown in **Fig. 7c**, it turns out that both in MLMs and GLMs, the trend of PCC is different from SCC which is extremely high in 2-fold change whereas R2, RMSE, and MAE are getting worse as the model size increases. As the fold change gets larger, the significance of the predicted values becomes more similar across the MLMs or GLMs ($p>0.05$). The results indicate that most pre-trained language models fail to predict high-frequency data accurately in the AMPCliff framework except for ESM2.

To investigate the capability of these LMs in general, we calculate the average values of the performance metrics of each MLM or GLM among all fold changes. As shown in **Fig. 7e**, it turns out that except for Recall and R2, all of the other 4 metrics get better as the model size of ESM2 increases. Whereas no metrics had the similar trend on GLMs (**Fig. 7f**). Note that all the R2 values are negative, indicating that the models are poor at predicting -log(MIC[M]) accurately.

We conclude the scaling law of ESM2 is available on the AMPCliff prediction task, while ProGen2 doesn't.

## Conclusions

This study identified and systematically defined the activity cliff (AC) phenomenon in AMPs (AMPCliff) as a variant of AC, and captured structurally similar peptides with significant differences in MIC values of antimicrobial activities. We leveraged the GRAMPA dataset to illustrate AMPCliffs with *Staphylococcus aureus* as a reference, and conducted a comprehensive benchmark analysis of predictive algorithms in this context.

Accurately quantifying peptide similarity proved essential, as existing sequence identity measures inadequately addressed short peptides. By developing a similarity calculation that considers the average alignment column similarity, we reframed peptide similarity as an amino-acid-level comparison. Our analysis of the BLOSUM62 substitution matrix is enriched with structural insights over traditional metrics like Tanimoto similarity, and underscores the need for a subtle similarity measure for non-canonical substructures. Additionally, our results suggest that current amino-acid-level models may benefit from greater structural detail within amino acid side chains, potentially enhancing models like ESM2 for future AMP prediction tasks.

We also introduced the AC split strategy, and its comparison against random split revealed that models trained on AC split performed better on detecting AMPCliffs with high-frequency signals based on the transferable knowledge learned from non-AC peptides. Our evaluation metrics, such as Recall and SCC, emphasized the predictive capacity across various ML and DL models. ESM2 demonstrated superior performance and scalability in AMPCliff prediction, and outperformed both fixed type and GLM-based learned representations, particularly with larger model sizes.

Our findings suggest that advanced pre-trained models with robust structural learning capabilities may hold the key to resolving AMPCliff challenges, and the results offer valuable insights for AMP design and potential applications in broader peptide-based therapeutic research.

## Materials and Methods

### Datasets

In 2018, Jacob Witten and Zack Witten introduced the GRAMPA dataset [57], which remains one of the only AMP datasets containing minimum inhibitory concentration (MIC) values. Previous studies have utilized GRAMPA for developing machine learning and deep learning models in AMP design [19,20], such as AMPSpace [19], CellFree-cnn [20], and CellFree-rnn [20], which we also employ here for comparative analysis on the AMPCliff prediction task.

To evaluate model performance, we implemented both random split and our proposed AC split strategies. Traditional data splitting methods commonly applied in small molecule drug design - such as scaffold, stratified, time, and target splits - are adapted to fit specific data characteristics. For example, time split captures historical trends in drug optimization [77], and scaffold split approximates these temporal patterns [78]. Target split [11] clusters data based on a shared target, fostering deep model learning of structural similarities, while avoiding out-of-distribution (OOD) issues associated with sequence diversity [11]. However, these methods are not ideally suited for antimicrobial peptide datasets like GRAMPA, which lacks time information and is highly heterogeneous, containing a variety of modified peptide structures absent in controlled corporate datasets.

Given these limitations, we developed the AC split method, inspired by previous work by Deng et al. [17] and Dablander et al. [1], which facilitates model learning from non-AC (low-frequency) to AC (high-frequency) peptides. In our implementation, AMPCliff pairs are progressively introduced to the training set by adjusting the fold-change threshold from 2 to 5, incrementally adding high-frequency data to evaluate model adaptability to AMPCliff predictions.

For comprehensive assessment, we employed both the AC split and a 5-fold stratified random split.

**Evaluation Metrics**

Model evaluation procedure can be substantially influenced by the choices of statistical analysis, metrics, and task settings [17]. We selected Spearman's rank correlation coefficient (SCC) and Recall as our primary evaluation metrics, prioritizing the relative ranking accuracy of predictions over their absolute label -log(MIC[M]) values. Additional metrics include R squared (R2), Root Mean Square Error (RMSE), Mean Absolute Error (MAE), and Pearson's correlation coefficient (PCC).

SCC quantifies the monotonic relationship between two variables (Eq. 6) by computing the correlation between their ranks, and its values range from -1 to 1. A value of ±1 signifies a perfect positively or inversely monotonic relationship. Unlike PCC, which assesses linear relationships and is highly sensitive to outliers, SCC is robust to outliers and accommodates non-linear trends, making it preferable when normality assumptions are unmet.

Recall is also known as sensitivity and reflects the proportion of positive samples correctly identified by the model. Here, Recall assesses the model's ability to identify peptides with strong antimicrobial properties by calculating the intersection of the top 50 ranked predicted and true values, based on their respective descending orders (Eq. 7), where larger label values -log(MIC[M]) indicate better antimicrobial activities.

RMSE and MAE are common error metrics in regression. RMSE measures the square root of the average squared differences between predictions and actual values, and is sensitive to outliers (Eq. 8). While MAE provides the mean absolute error, and reflects average deviations from observed values (Eq. 9).

PCC (Eq. 10) evaluates the linear correlation between predictions and labels, with values ranging from -1 to 1. A value of ±1 indicates a perfect linear relationship, while zero implies no linear correlation.

The coefficient of determination (R2, Eq. 11) measures the proportion of variance in predictions explained by observed values, and it generally ranges from 0 to 1. An R2 of 1 reflects an ideal fit, while values closer to 0 indicate limited or poor predictive power compared to the mean-based baseline.

$$SCC = \frac{\sum_{i=1}^{N}\left(Rank(y_i) - \overline{Rank}(y_{obs})\right)\left(Rank(\hat{y}_i) - \overline{Rank}(y_{pred})\right)}{\sqrt{\sum_{i=1}^{N}\left(Rank(y_i) - \overline{Rank}(y_{obs})\right)^2 \sum_{i=1}^{N}\left(Rank(\hat{y}_i) - \overline{Rank}(y_{pred})\right)^2}} \quad (6.)$$

$$Recall = Rank_K(y_i)[:K] \bigcap Rank_K(y_{pred})[:K], K = 50 \quad (7.)$$

$$RMSE = \sqrt{\frac{1}{N}\sum_{i=1}^{N}(y_i - \hat{y}_i)^2} \quad (8.)$$

$$MAE = \frac{1}{N}\sum_{i=1}^{N}|y_i - \hat{y}_i| \quad (9.)$$

$$PCC = \frac{\sum_{i=1}^{N}(y_i - \bar{y}_{obs})(\hat{y}_i - \bar{y}_{pred})}{\sqrt{\sum_{i=1}^{N}(y_i - \bar{y}_{obs})^2 \sum_{i=1}^{N}(\hat{y}_i - \bar{y}_{pred})^2}} \quad (10.)$$

$$R2 = 1 - \frac{\sum_{i=1}^{N}(y_i - \hat{y}_i)^2}{\sum_{i=1}^{N}(y_i - \bar{y}_{obs})^2} \quad (11.)$$

## Model Training

This study used default hyperparameter values for ML models. For DL models, we followed hyperparameter values reported in the literature. Specifically, following [19], we configured the LSTM with a hidden dimension of 128, embedding dimension of 50, dropout of 0.7, 2 layers, and unidirectional mode. For CNN and RNN models, we set the CNN kernel size to 5, with 64 output channels across 2 layers, followed by 3 fully connected layers [20]. The RNN (LSTM) model used a hidden dimension of 500 and was followed by a single fully connected layer. Following [40], we set the fingerprint radius to 3 and bit count to 2048, with a CNN of kernel size 2, 256 output channels across 3 layers, and a final fully connected layer. Detailed configurations are in Supplementary Table S1. For language models, we retained the default hyperparameters as outlined in Supplementary Table S5.

All experiments with neural network and language models were conducted on a single NVIDIA V100 GPU for 100 epochs and 50 epochs, respectively. The validation loss guided the selection of the best model during training for final testing. Batch size was set to 4, as smaller batches facilitate learning of the refined, localized features characteristic of AC, while larger batches typically aid in capturing broader and general features. To ensure a fair comparison, we saved all raw predictions, and evaluation metrics were consistently computed using identical code.

**Statistical analysis**

To assess statistically significant differences among models and representations, we employed methods consistent with prior work (Deng et al., 2023), and conducted comprehensive statistical analyses on prediction performance (Supplementary Tables 9-11). Specifically, we utilized the non-parametric Wilcoxon rank-sum test to account for non-normal data distributions. A significance threshold of two-sided $p<0.05$ was applied throughout.

# Acknowledgements

This work was supported by the Senior and Junior Technological Innovation Team (20210509055RQ), the National Natural Science Foundation of China (62072212 and U19A2061), Guizhou Provincial Science and Technology Projects (ZK2023-297), the Science and Technology Foundation of Health Commission of Guizhou Province (gzwkj2023-565), Science and Technology Project of Education Department of Jilin Province (JJKH20220245KJ and JJKH20220226SK), the Jilin Provincial Key Laboratory of Big Data Intelligent Computing (20180622002JC), the Development Project of Jilin Province of China (No. 20220508125RC), and the Fundamental Research Funds for the Central Universities, JLU.

# Code and Data availability

The raw datasets and the source code of AMPCliff are available at https://github.com/Kewei2023/AMPCliff-generation or https://www.healthinformaticslab.org/supp/.